\documentclass[aps,prd,preprint,a4paper]{revtex4} 
\usepackage{epsfig}
\begin{document}
\title{On the Einstein-Cartan cosmology vs. Planck data}
\author{Davor Palle}
\affiliation{
Zavod za teorijsku fiziku, Institut Rugjer Bo\v skovi\' c \\
Bijeni\v cka cesta 54, 10000 Zagreb, Croatia}

\date{September 20, 2013}

\begin{abstract}
The first comprehensive analyses of Planck data reveal that the cosmological
model with dark energy and cold dark matter can satisfactorily explain the essential
physical features of the expanding Universe.
However, the inability to simultaneously fit large and small
scale TT power spectrum, scalar power index smaller than one and the observations of
the violation of the isotropy found by few statistical indicators of the CMB, 
urge theorists to search for explanations.
We show that the model of the Einstein-Cartan cosmology with clustered dark matter halos and 
their corresponding clustered angular momenta coupled to torsion, can account
for small scale - large scale discrepancy and larger peculiar velocities (bulk flows) for
galaxy clusters.
The nonvanishing total angular momentum (torsion) of the Universe enters as a negative effective
density term in the Einstein-Cartan equations causing partial cancellation of
the mass density.
The integrated Sachs-Wolfe contribution of the Einstein-Cartan model is negative, thus it can
provide partial cancellation of the large
scale power of the TT CMB spectrum.
The observed violation of the isotropy appears as a natural ingredient of the Einstein-Cartan 
model caused by the spin densities of light Majorana neutrinos in the early stage of the
evolution of the Universe and bound to the lepton CP violation and matter-antimatter
asymmetry.
\end{abstract}

\maketitle 

\vspace{2ex}

\section{Introduction and motivation}

Although the presence of dark matter and dark energy is justified by all
cosmological observations, their identification and properties are still
far from being established. The measurements of the fluctuations of the CMB
are in this respect escpecially valuable because of the wealth and accurate
information that can be extracted from them.

The most recent disclosed results of the Planck mission contain issues like:
the temperature power spectrum, gravitational lensing or integrated Sachs-Wolfe
effect, up to Sunyaev-Zeldovich cluster counts, isotropy,
nonGaussianity or cosmic infrared background.
It seems that the old, unexpected features, beyond the $\Lambda CDM$ + inflation
model, persist in data and are even more highlighted: 1. large scale temperature
power spectrum much lower than the $\Lambda CDM$ prediction, not limited only
to low quadrupole \cite{WMAP} but to almost all multipole moments $l < 30$ (see Fig. 37 in
ref. \cite{Planck01}), 2. scalar power spectrum index less than 1 (see Table 8
in ref. \cite{Planck01}), 3. violation of isotropy observed as hemispherical
asymmetry, parity asymmetry, quadrupole-octopole alignment, cold spot, and dipolar
asymmetry \cite{Planck02}.

If the violation of isotropy will be confirmed by other complementary cosmic
observations of radio galaxies \cite{Birch},
spiral galaxies \cite{Longo}, bulk flows of clusters \cite{Kashlinsky} or quasars \cite{Hutsemekers},
it will challenge cosmological principles and call for the new theoretical 
insights.

Assuming that the observed anomalies are real phenomena, we try to understand and elucidate
measured physical features by the Einstein-Cartan (EC) cosmology. 
Incorporating rotating degrees of freedom of matter (spin and angular momentum)
and spacetime (torsion) into the relativistic framework, the EC cosmology appears as a 
nonsingular theory \cite{Trautman,Palle01}, the cosmic mass density can be fixed \cite{Palle01},
the scalar power index can acquire the negative tilt \cite{Palle01a},
 spin densities triger density fluctuations \cite{Palle02}
and the right-handed vorticity (rotation) of the Universe \cite{Palle03} resulting at later stages of
the evolution in the nonvanishing total angular momentum of the Universe \cite{Palle04}.
The nonsingular EC cosmology is in conformity with the nonsingular theory
of gauge interactions in particle physics \cite{Palle05} that contains light and heavy
Majorana neutrinos as hot and cold dark matter particles \cite{Palle06}, including
other important implications of the perturbative and nonperturbative aspects of strong and
electroweak interactions phenomenology \cite{Palle07}.

In this paper we investigate and compare EC and $\Lambda CDM$ cosmologies solving
evolution equations for the
scale dependent density contrasts, mass fluctuations, peculiar velocities, and
integrated Sachs-Wolfe effect. The next chapter contains description of the
evolution equations, definitions and the introduction of our simple clustering model.
The concluding section deals with the numerical results of the computations,
comparisons of the EC, and $\Lambda CDM$ cosmologies and final remarks and hints
for future research.		

\section{Definitions, equations and clustering model}

Because any deviation from cosmic homogeneity and isotropy is very small, we limit our
considerations to the homogeneous and isotropic geometry. We start the evolution
in the radiation era when clustering of dark and baryonic matter is negligible.
The evolution equations for matter density contrasts in Fourier space
are derived in \cite{Peebles}

\begin{eqnarray}
\frac{d^{2}h}{d t^{2}}&+&\frac{2}{a}\frac{d a}{d t}\frac{d h}{d t} = 8\pi G_{N}(2\rho_{r}
\delta_{r} + \rho_{m}\delta_{m}),  \nonumber \\
\frac{d \delta_{m}}{d t} &=& \frac{1}{2}\frac{d h}{d t},\ 
\frac{d \delta_{r}}{d t} = \frac{4}{3} (\frac{k v }{a} +\frac{1}{2}\frac{d h}{d t}), \nonumber \\
\frac{d v}{d t} &=& -k \frac{\delta_{r}}{4 a},\ 
(\frac{d a}{d t})^{2}=\frac{8}{3}\pi G_{N} a^{2}(\rho_{r}+\rho_{m}+\rho_{\Lambda}).
\end{eqnarray}

We here use the notation
 $h=\sum_{\alpha=1}^{3}h_{\alpha}^{\alpha}, g_{\alpha\beta}=-a^{2}[\delta_{\alpha\beta}-
h_{\alpha\beta}]$, density contrasts are $\delta_{i}=\frac{\delta \rho_{i}}{\rho_{i}}$,
k is the comoving wave number, $a=\frac{R}{R_{0}}=\frac{1}{1+z}$,
subscripts m,r,$\Lambda$ denote matter, radiation and the cosmological constant quantities,
respectively, and
v is a velocity. All the quantites are functions of t and $\vec{k}$.

These equations can be cast into the more suitable form by getting rid of 
$h=\sum_{\alpha=1}^{3}h_{\alpha}^{\alpha}$
and by changing the evolution variable to $y=\ln a$:

\begin{eqnarray}
\frac{d^{2}\delta_{m}}{d y^{2}} &=& -\frac{1}{2}\frac{d \delta_{m}}{d y} \Omega_{m}
(\Omega_{r}a^{-1}+\Omega_{m}+\Omega_{\Lambda}a^{3})^{-1}+\frac{3}{2}(2 \Omega_{r}\delta_{r}
+\Omega_{m} a \delta_{m})(\Omega_{r}+\Omega_{m}a+\Omega_{\Lambda}a^{4})^{-1}, \nonumber \\
\frac{d \delta_{r}}{d y} &=& \frac{4}{3}(\frac{d \delta_{m}}{d y}+\frac{k v }{\dot{a}}),
\ \frac{d v}{d y} = -\frac{\delta_{r}}{4}\frac{k}{\dot{a}}. \hspace{50 mm}
\end{eqnarray}

Our notation includes $\rho_{r}=\Omega_{r}\rho_{c}a^{-4}$, $\rho_{m}=\Omega_{m}\rho_{c}a^{-3}$,
 $\rho_{\Lambda}=\Omega_{\Lambda}\rho_{c}$, $\rho_{c}=\frac{3 H_{0}^{2}}{8\pi G_{N}}$,
$H_{0}=100 h\ km s^{-1}Mpc^{-1}$ and $\dot{a}=3.2409\times 10^{-18} h [\Omega_{r}a^{-2}+\Omega_{m}a^{-1}
+\Omega_{\Lambda}a^{2}]^{1/2} s^{-1}$.

The evolution equations for the EC cosmology (neglecting small vorticity and acceleration $\omega=m=0,\  
\lambda=0;\ Q=Q_{0}a^{-3/2}=torsion$) are derived in \cite{Palle04} (eq. (14)):

\begin{eqnarray}
\ddot{\delta}_{1}&+&2\frac{\dot{a}}{a}\dot{\delta}_{1}-2 Q \dot{\delta}_{2}+
(-\frac{1}{3}\kappa \Lambda -\frac{5}{3}Q^{2}-\frac{1}{3}\kappa \rho +\frac{\ddot{a}}{a})\delta_{1}
+\frac{1}{4}\frac{\dot{a}}{a}Q \delta_{2} = 0 , \nonumber \\
\ddot{\delta}_{2}&+&2\frac{\dot{a}}{a}\dot{\delta}_{2}+2 Q \dot{\delta}_{1}+
(-\frac{1}{3}\kappa \Lambda -\frac{5}{3}Q^{2}-\frac{1}{3}\kappa \rho +\frac{\ddot{a}}{a})\delta_{2}
-\frac{1}{4}\frac{\dot{a}}{a}Q \delta_{1} = 0 , \nonumber \\
\ddot{\delta}_{3}&+&2\frac{\dot{a}}{a}\dot{\delta}_{3}+
(-\frac{1}{3}\kappa \Lambda -\frac{2}{3}Q^{2}-\frac{1}{3}\kappa \rho +\frac{\ddot{a}}{a})\delta_{3} = 0, \\
\delta &\equiv& [\delta_{1}^{2}+\delta_{2}^{2}+\delta_{3}^{2}]^{1/2}. \nonumber
\end{eqnarray}

We assume that after redshift $z_{G} = 10$ the nonlinear bound structures are formed  
in the form of stars, galaxies, and clusters. The clustering of particles forming halos are
described by a model with only two parameters $k_{G}$ and $\sigma_{G}$. This is applied to both mass and angular
momentum clustering:

\begin{eqnarray}
Q(a)=(2\pi)^{-3}\int d^{3}k \hat{Q}(a,\vec{k})=Q_{0} a^{-3/2}\Theta(z_{G}-z), \nonumber \\
\hat{Q}(a,\vec{k})=\bar{Q}_{0}a^{-3/2}e^{-\mid k-k_{G} \mid /\sigma_{G}}\Theta(z_{G}-z)
\Rightarrow Q_{0}=\bar{Q}_{0}(2 \pi)^{-3} \int d^{3} k e^{-\mid k-k_{G} \mid /\sigma_{G}}, \\
\rho(a)=(2\pi)^{-3}\int d^{3}k \hat{\rho}(a,\vec{k})=\rho_{0} a^{-3}\Theta(z_{G}-z), \nonumber \\
\hat{\rho}(a,\vec{k})=\bar{\rho}_{0}a^{-3}e^{-\mid k-k_{G} \mid /\sigma_{G}}\Theta(z_{G}-z)
\Rightarrow \rho_{0}=\bar{\rho}_{0}(2 \pi)^{-3} \int d^{3} k e^{-\mid k-k_{G} \mid /\sigma_{G}}, 
\end{eqnarray}

Fourier transforms of the evolution equations (3) take now the following form with the 
evolution variable $y=\ln a$:

\begin{eqnarray}
(\frac{\dot{a}}{a})^{2}\frac{d^{2}\Delta_{1,2}}{d y^{2}}+\frac{\dot{a}}{a}(\frac{d\dot{a}}{d a}
+\frac{\dot{a}}{a})\frac{d \Delta_{1,2}}{d y} \mp 2\frac{\dot{a}}{a}\langle Q\frac{d \delta_{2,1}}{d y}
\rangle_{FT} + (-\frac{1}{3}\kappa \Lambda+\frac{\ddot{a}}{a})\Delta_{1,2}  \nonumber \\
-\frac{1}{3}\kappa \langle \rho_{m}\delta_{1,2} \rangle_{FT}
-\frac{5}{3} \langle Q^{2}\delta_{1,2} \rangle_{FT}
\pm\frac{1}{4}\frac{\dot{a}}{a} \langle Q\delta_{2,1} \rangle_{FT} = 0, \nonumber \\
(\frac{\dot{a}}{a})^{2}\frac{d^{2}\Delta_{3}}{d y^{2}}+\frac{\dot{a}}{a}(\frac{d\dot{a}}{d a}
+\frac{\dot{a}}{a})\frac{d \Delta_{3}}{d y} + (-\frac{1}{3}\kappa\Lambda+\frac{\ddot{a}}{a})\Delta_{3} 
-\frac{1}{3}\kappa \langle \rho_{m}\delta_{3} \rangle_{FT} -\frac{2}{3}\langle Q^{2}\delta_{3}\rangle_{FT}=0.
\end{eqnarray}

The Einstein-Cartan field equations define the cosmic clocks (see eq. (15) of \cite{Palle04})
as follows:

\begin{eqnarray}
\dot{a}=H_{0}[\Omega_{m}a^{-1}+\Omega_{\Lambda}a^{2}-\frac{1}{3}a^{2}Q^{2}]^{1/2}, \nonumber \\
\frac{d\dot{a}}{d a}=H_{0}\frac{1}{2}[\Omega_{m}a^{-1}+\Omega_{\Lambda}a^{2}-\frac{1}{3}a^{2}Q^{2}]^{-1/2}
[-\Omega_{m}a^{-2}+2 \Omega_{\Lambda}a+\frac{1}{3}aQ^{2}], \nonumber \\
\frac{\ddot{a}}{a}=\frac{1}{3}\kappa \Lambda -\frac{1}{6}\kappa \rho +\frac{2}{3}Q^{2}.
\end{eqnarray}

The following definitions and convolutions are used in eq. (6):

\begin{eqnarray*}
\Delta_{i}(y,\vec{k}) &\equiv& \int d^{3}x e^{\imath \vec{k}\cdot\vec{x}}\delta_{i}(y,\vec{x}), \\
\langle Q \delta_{i} \rangle_{FT}(y,\vec{k}) &\equiv& \int d^{3}x e^{\imath \vec{k}\cdot\vec{x}}
Q(y,\vec{x})\delta_{i}(y,\vec{x})=(2\pi)^{-3}\int d^{3} k' Q(y,\vec{k}-\vec{k}')\Delta_{i}(y,\vec{k}'), \\
\langle Q^{2} \delta_{i} \rangle_{FT}(y,\vec{k}) &\equiv& \int d^{3}x e^{\imath \vec{k}\cdot\vec{x}}
Q^{2}(y,\vec{x})\delta_{i}(y,\vec{x}) \\
&=& (2\pi)^{-6}\int d^{3} k'd^{3} k''\Delta_{i}(y,\vec{k}')Q(y,\vec{k}'')Q(y,\vec{k}-\vec{k}'-\vec{k}'').
\end{eqnarray*}

Having all the evolution equations for the EC and $\Lambda CDM$ cosmologies, we define initial conditions
in the radiation era and choose the parameters of the models:

\begin{eqnarray*}
a_{i}=10^{-8},\ \delta_{r}(a_{i})=k^{1/2}a_{i}^{2},\ \delta_{m}(a_{i})=\frac{3}{4}k^{1/2}a_{i}^{2},\ y=\ln a, \\
\frac{d \delta_{r}}{d y}(a_{i})=2 k^{1/2}a_{i}^{2},\ 
\frac{d \delta_{m}}{d y}(a_{i})=\frac{3}{2}k^{1/2}a_{i}^{2},\ v(a_{i})=0, \\
\Lambda CDM: \Omega_{m}=0.34,\ \Omega_{\Lambda}=0.66,\ Q=0,\ h=0.67, \\
EC: \Omega_{m}=2,\ \Omega_{\Lambda}=0,\ h=0.67,\ Q=\left\{
\begin{array}{ll}
0 & if\ z > 10 \\
-2.3 a^{-3/2} & if\ 1 < z \leq 10 \\
-\sqrt{3} a^{-3/2} & if\  0 \leq z \leq 1
\end{array}
\right. 
\end{eqnarray*}

Our choice of the torsion (angular momentum) parameters is guided by the condition that at zero redshift
$\Omega_{Q} \simeq -1$ \cite{Palle01,Palle04} (at the redshifts $1 \geq z \geq 0$ 
the galaxy clusters emerge changing the total angular momentum contribution of the era $z > 1$),
while at the earlier epoch $10 > z > 1$ 
our choice is guided by the condition to match roughly the
correct cosmic clocks and the age of the Universe:

\begin{eqnarray*}
\tau_{U}(Gyr)=\frac{10}{h}\int^{1}_{10^{-3}} \frac{d a}{a}[\Omega_{\Lambda}
+\Omega_{m}a^{-3}-\frac{1}{3}Q^{2}]^{-1/2} , \\
\tau_{U}(\Lambda CDM)=13.89 Gyr,\ \tau_{U}(EC)=13.29 Gyr, \\
k_{min}=10^{-3} Mpc^{-1},\ k_{max}=10^{2} Mpc^{-1},
\ k_{G}=1 Mpc^{-1},\ \sigma_{G}=0.25 Mpc^{-1}.
\end{eqnarray*}

We integrate the above evolution equations to the relative accuracy ${\cal O}(10^{-4})$ by lowering the
integration steps until the required accuracy is reached.
Equations (2) are solved for the evolution from $a_{i}=10^{-8}$ up to $a_{G}=1/(1+z_{G})$ and Eqs. (6)
are then solved from $a_{G}=1/(1+z_{G})$ up to $a=1$.
The Adams-Bashforth-Moulton predictor-corrector method is used for differential equation integrations
(code of L. F. Shampine and M. K. Gordon, Sandia Laboratories, Albuquerque, New Mexico)
and CUBA library for multidimensional integrations \cite{Hahn}.
The next section is devoted to the detailed exposure of the numerical results and
comparison between EC and $\Lambda CDM$ models. The relevance of the results for the Planck data are
given here as well.

\section{Results, discussion and conclusions}

Because the best fit to the Planck temperature power spectrum is dominantly performed by the multipoles
$l > 30$, it is not a surprise that the concordance $\Lambda CDM$ model is favoured,
but at the expense of the wrong fit for low mutipoles (large scales). By solving evolution equations
for the EC and $\Lambda CDM$ with the simple clustering model, one can verify that at low redshifts
these two models produce density contrasts that differ substantially at large scales, while being
similar at smaller scales (see Fig. 1).

\epsfig{figure=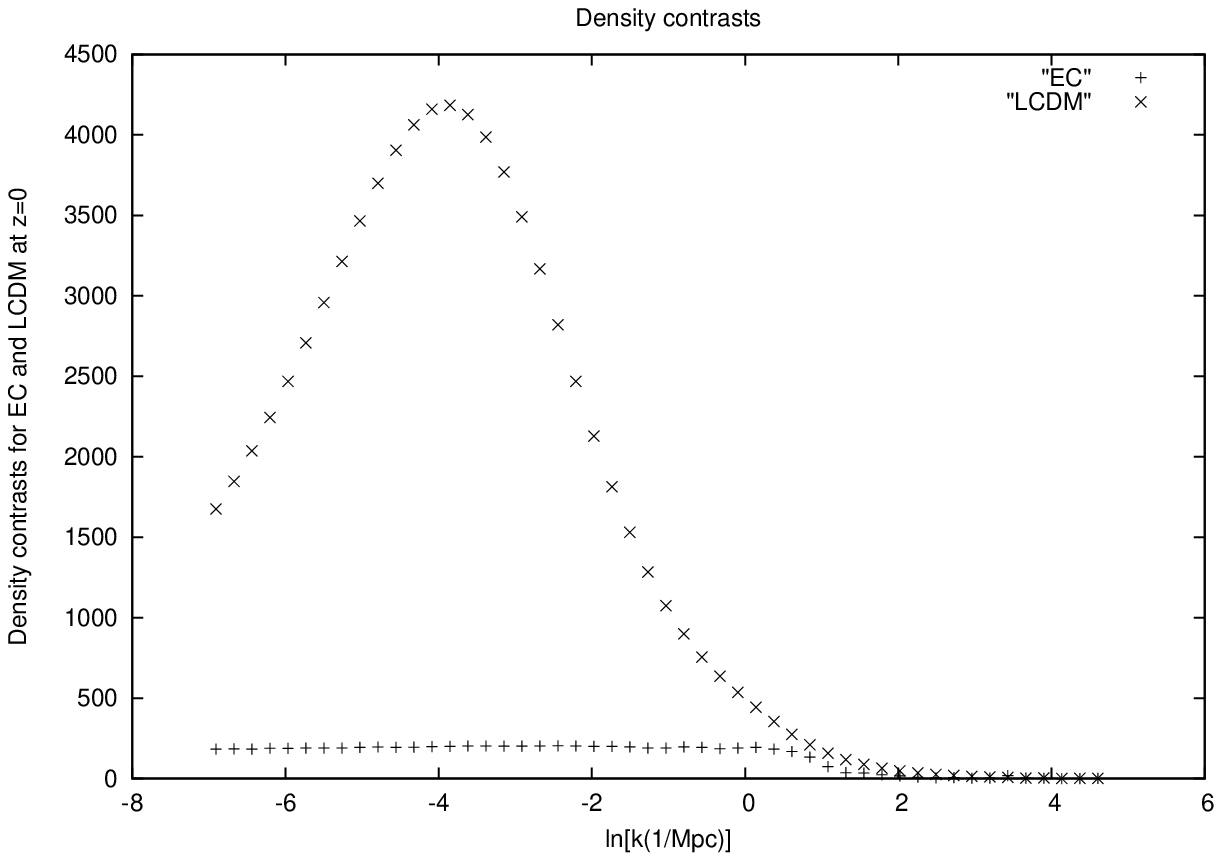, height=90 mm, width=130 mm}

\vspace{2mm}

{\bf Fig. 1: Density contrasts at z=0 as functions of the wave number $k(Mpc^{-1})$
 normalized at $k_{max}: \delta_{m}(k_{max})=1$.
}
\newline

If we accept the following normalization on a larger scale \cite{Palle04}:

\begin{eqnarray*}
(\delta M/M)_{RMS}(a=1,S_{0}=10 h^{-1}Mpc) = 1 .
\end{eqnarray*}

the processed spectrum of the mass fluctuations with the top hat window function for EC and $\Lambda CDM$ models
then differ at small scales (see Fig. 2):

\begin{eqnarray}
(\delta M/M)^{2}_{RMS}(a,S) \equiv
 N^{-1} \int d^{3}k W^{2}(\vec{k},S)
 |\delta (a,\vec{k})|^{2}, \\
N=\int d^{3}k W^{2}(\vec{k},S_{0})|\delta (a=1,\vec{k})|^{2}, \nonumber \\
W(y=kS)=\frac{3}{y^{3}}(\sin y-y\cos y). \nonumber
\end{eqnarray}

\epsfig{figure=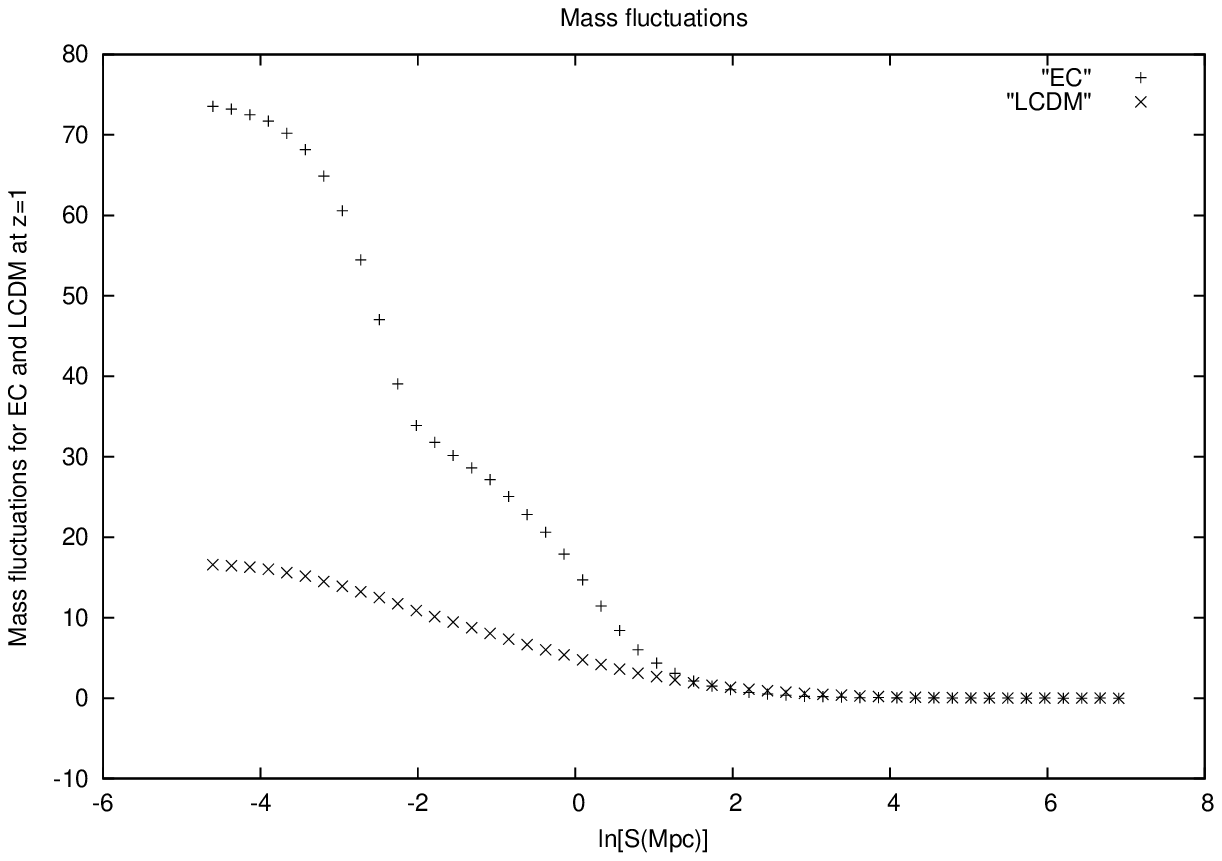, height=90 mm, width=130 mm}
\epsfig{figure=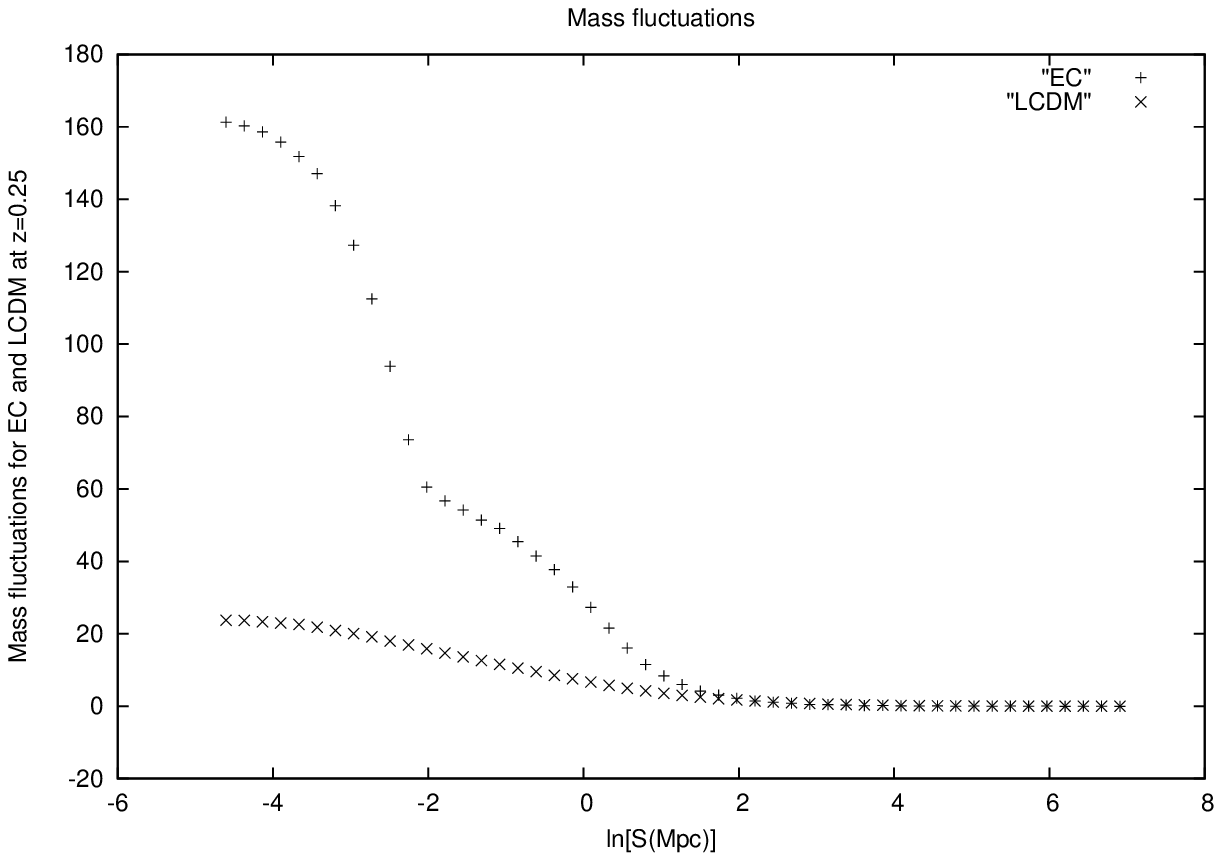, height=90 mm, width=130 mm}
\epsfig{figure=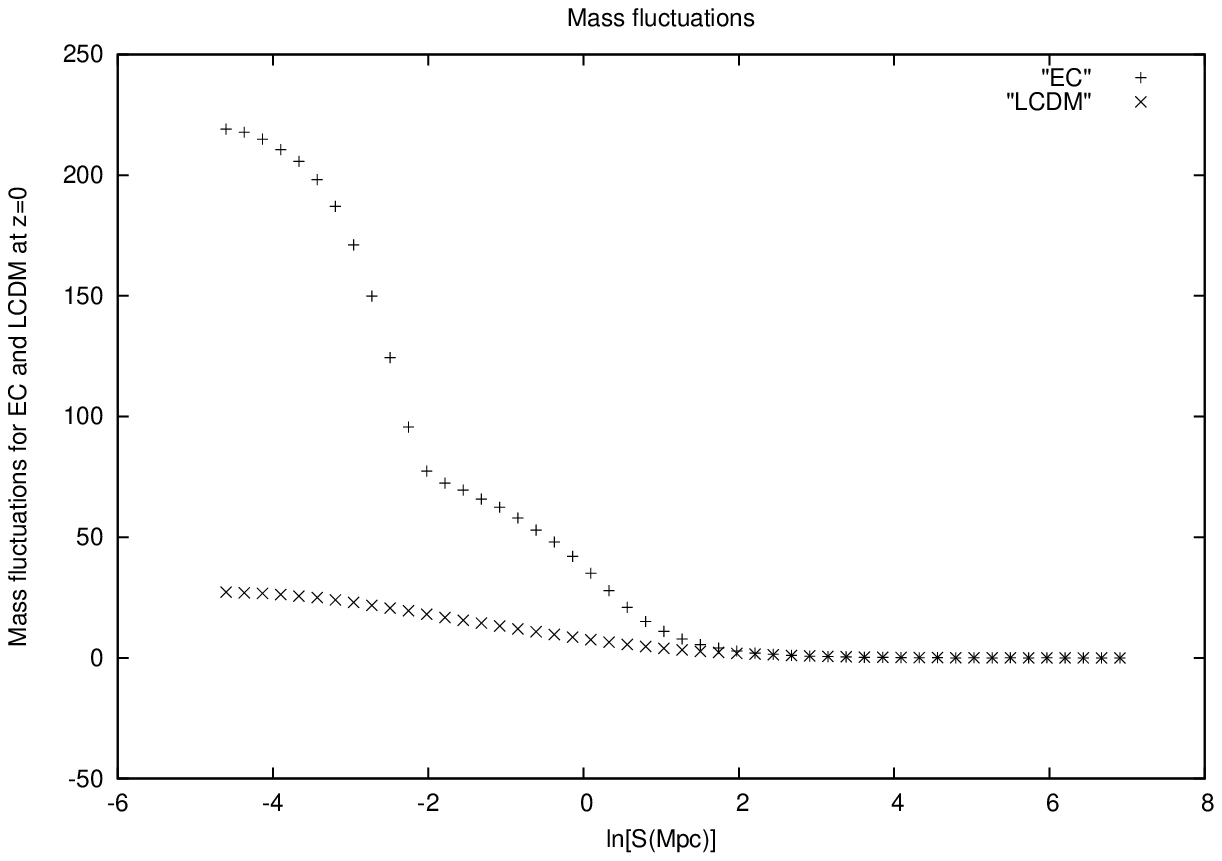, height=90 mm, width=130 mm}

\vspace{2mm}

{\bf Fig. 2: Mass fluctuations for three redshifts (z=0;0.25;1) as 
functions of the scale S(Mpc).
}
\newline

We can similarly evaluate the peculiar velocities as functions of the scale and redshift with
the same normalization as in eq.(8):

\begin{eqnarray}
v_{RMS}^{2}(a,S) \equiv  N^{-1} 
\int d^{3}k W^{2}(\vec{k},S) \frac{1}{\vec{k}^2}
|a \dot{a}\frac{d \delta (a,\vec{k})}{d a}|^{2} , 
\end{eqnarray}

giving the expected results (see Fig. 3), where the EC cosmology produces larger peculiar velocities 
than the $\Lambda CDM$ cosmology at the galaxy and galaxy cluster scales 
${\cal O}(10^{-1}) Mpc-{\cal O}(10^{2}) Mpc$.

\epsfig{figure=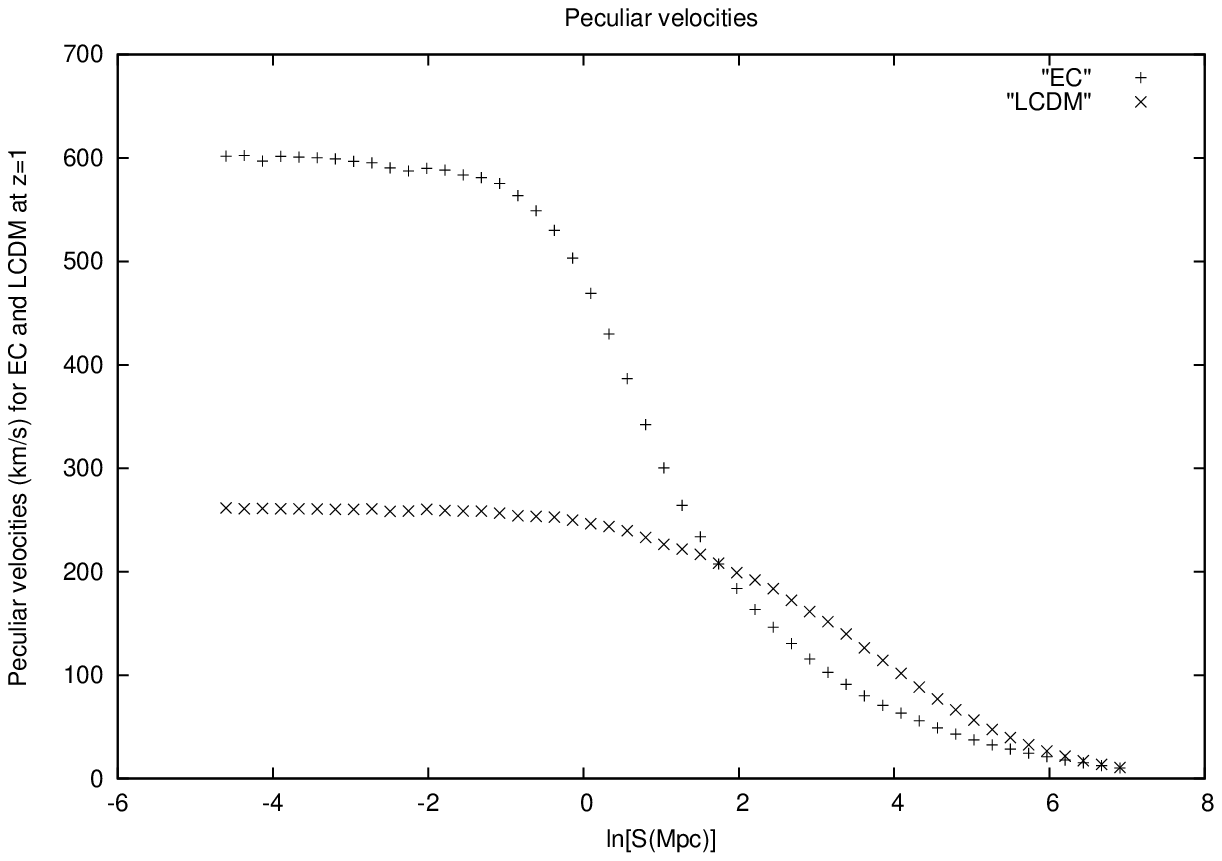, height=90 mm, width=130 mm}
\epsfig{figure=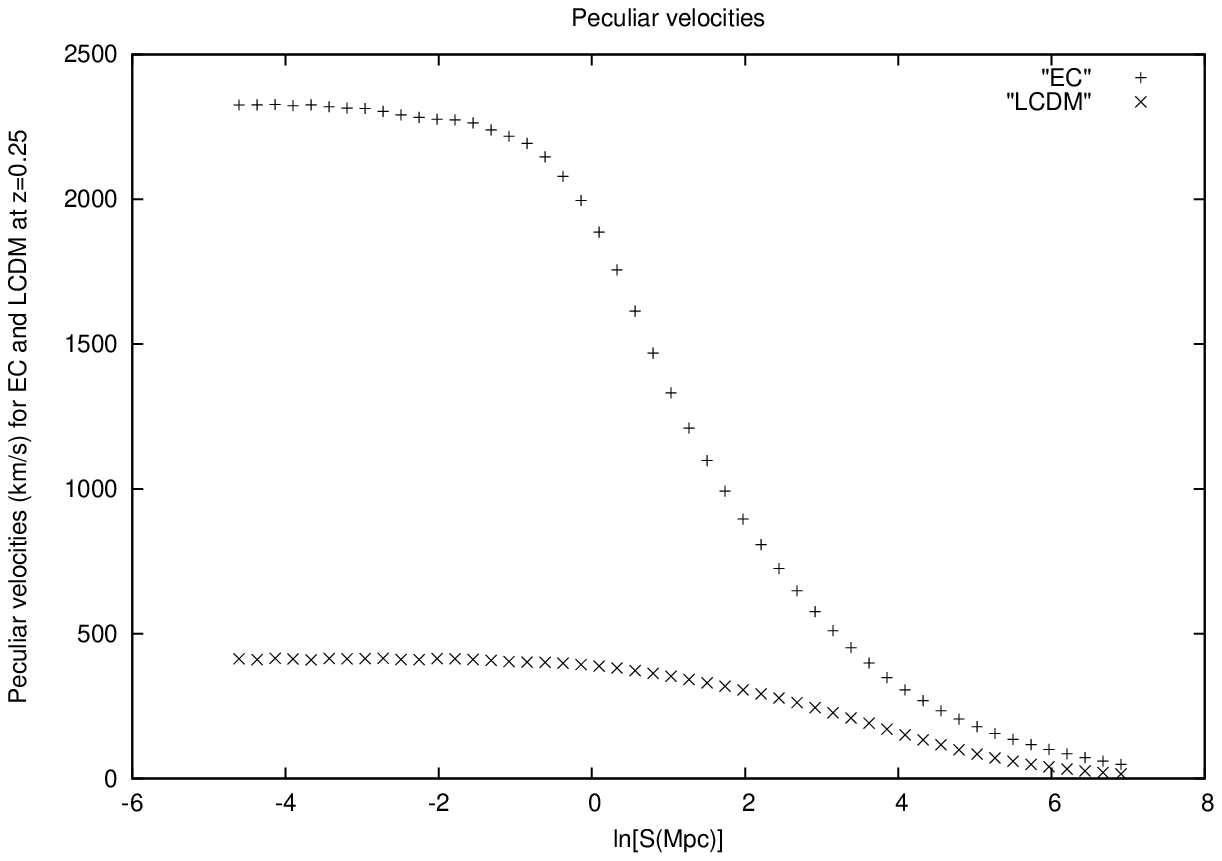, height=90 mm, width=130 mm}
\epsfig{figure=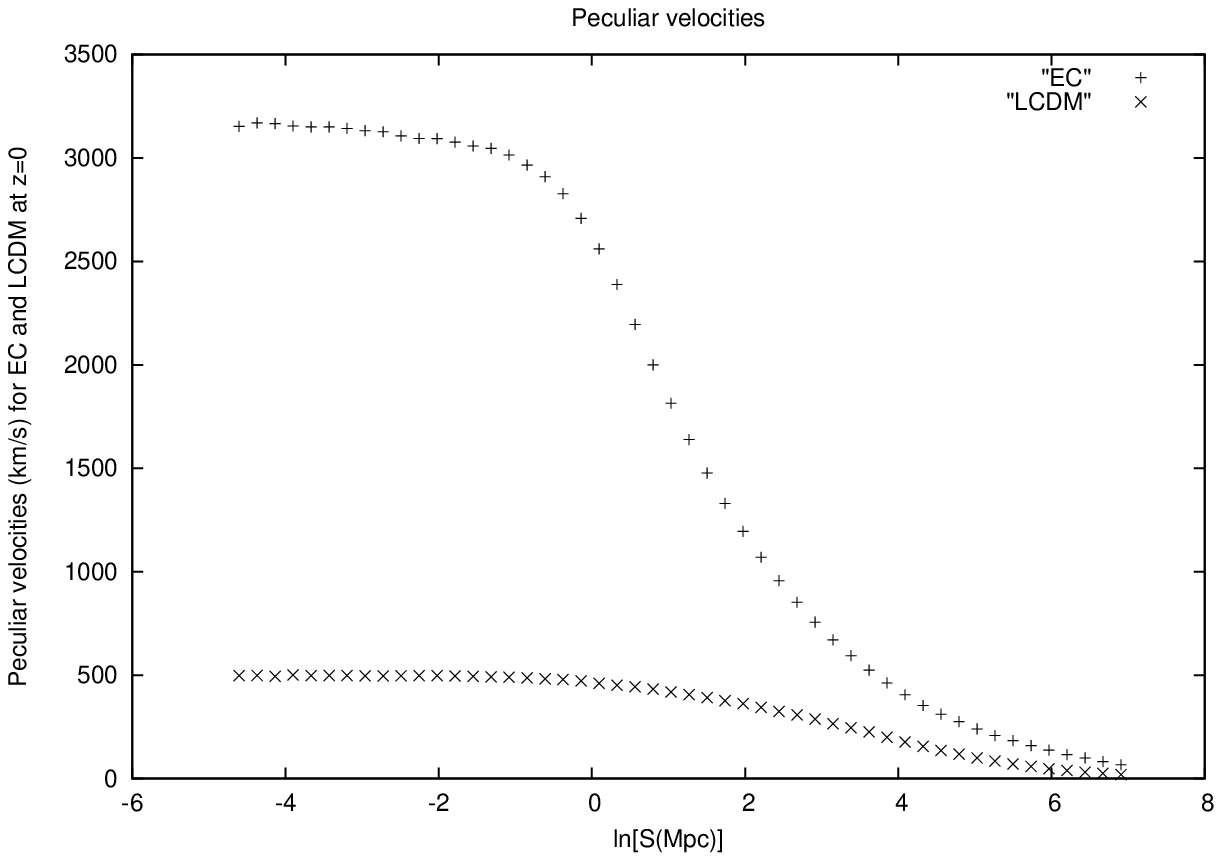, height=90 mm, width=130 mm}

\vspace{2mm}

{\bf Fig. 3: Peculiar velocities for three redshifts (z=0;0.25;1) as 
functions of the scale S(Mpc).
}
\newline

The conclusions are not sensitive to the reasonable choices of the parameters of the clustering model.
The integrated Sachs-Wolfe (ISW) effect plays an important role at low redshifts in the evolution, 
if the mass density differs from one \cite{Palle04}:

\begin{eqnarray}
a_{lm}^{ISW} = 12\pi \imath^{l}\int d^{3}k Y^{m*}_{l}(\hat{k})
\delta_{\vec{k}} (\frac{H_{0}}{k})^{2}
\int da j_{l}(kr)\chi^{ISW}, \\
\chi^{ISW}(a)=-\Omega_{m}\frac{d}{da} (\delta (a)/a),
\ r=\int^{1}_{a} da a^{-2}H^{-1}(a),\ \delta (a=1)=1. \nonumber
\end{eqnarray}

The ISW is positive (negative) for the $\Lambda CDM$ (EC) cosmology (see Fig. 4). 
The structure of the EC ISW curve around z=1 is just an artefact of our simple model for torsion with
a nonanalytic behaviour at z=1.

\epsfig{figure=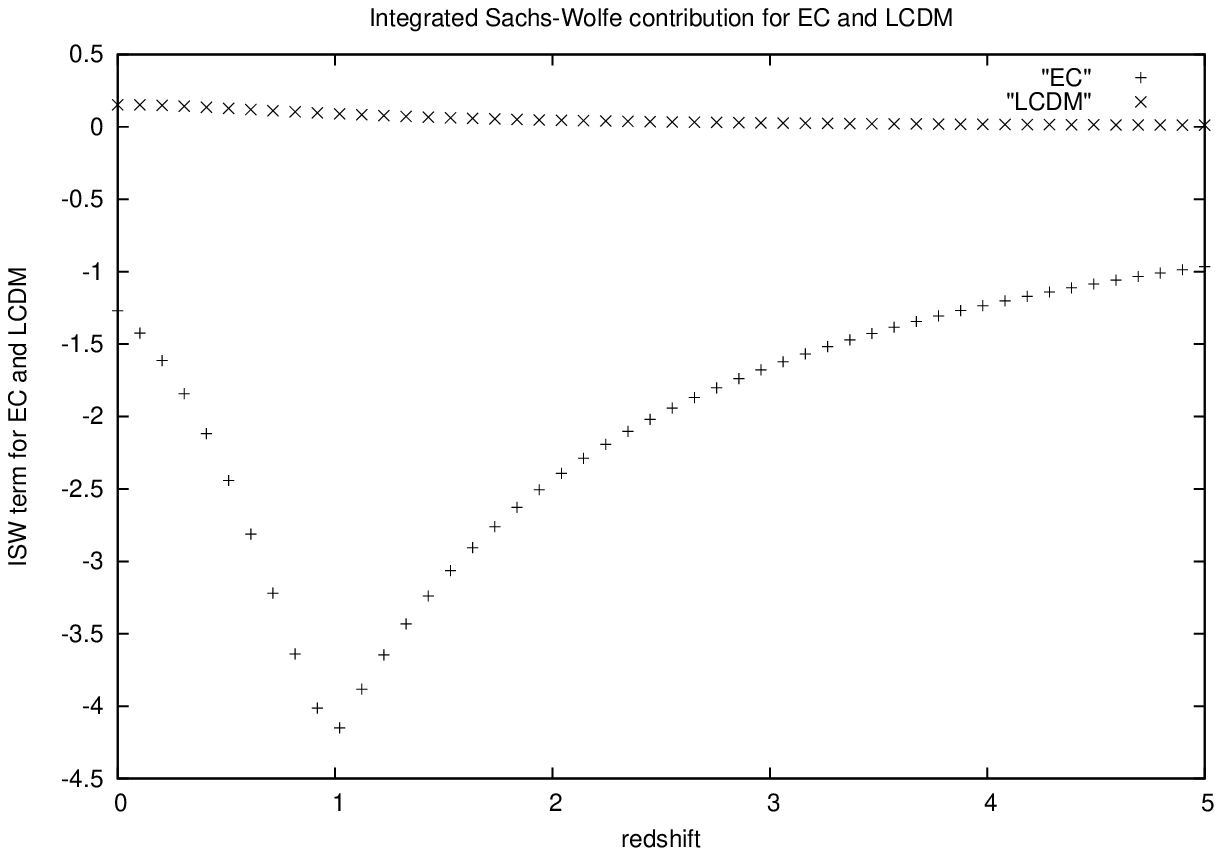, height=90 mm, width=130 mm}

\vspace{2mm}

{\bf Fig. 4: Integrated Sachs-Wolfe terms $\chi^{ISW}$ for
EC and $\Lambda$CDM cosmologies.
}
\newline

The compendium of all our results can be summarized as follows: (1) the $\Lambda CDM$ model cannot  
simultaneously fit the large and the small scale parts of the Planck TT spectrum, while the EC can rectify
this deficiency owing to the presence of the new rotational degrees of freedom (torsion)
that partially cancels the large mass density ($\Omega_{m}=2$) when clustering matters, i.e. the rotation
(centripetal force) acts opposite to the attractive force of gravity, (2) the presence of the ISW effect is observed in 
Planck data \cite{Planck03}, but with the unknown sign; the very small low multipoles of the Planck
TT spectrum imply the negative contribution of the ISW \cite{Kodama}, thus in accord with the EC model,
(3) the peculiar velocities are larger at the galaxy and galaxy cluster scales for the EC than $\Lambda CDM$ cosmologies
at low redshifts.
These conclusions are robust and are qualitatively valid for the reasonable variation of the clustering model parameters
$k_{G}$ and $\sigma_{G}$.
The two different
analyses of the Planck peculiar velocities of galaxy clusters \cite{Planck04} are still not conclusive 
as to whether the data are consistent with the $\Lambda CDM$ model or not. 

Our final remark is that the first Planck results favour a description of the Universe with the
anisotropic models.
 The $\Lambda CDM$ and the inflationary paradigm cannot fulfil
severe phenomenological requirements. We show that the EC cosmology with the new rotational degrees of
freedom can resolve almost all of the $\Lambda CDM$ model deficiencies. However, the N-body numerical simulations, 
 including the angular momenta of the CDM halos and their feedback
onto the background cosmic geometry, have to be applied within the EC gravity.

\end{document}